\documentclass[aip,jmp,reprint,amsmath,amssymb]{revtex4-1}

\usepackage{graphicx}
\usepackage{epsfig}
\usepackage{dcolumn}
\usepackage{bm}
\newcommand     {\beq}[1]         { \begin{equation} #1 \end{equation} }

\begin{document}

\title{Statistical features of magnetic noise in mixed-type impact fracture} 

\author{Zs.\ Danku$^{1}$, Gy.\ B.\ Lenkey$^{2}$, 
and F.\ Kun$^{1}$\email{ferenc.kun@science.unideb.hu}}
\address{
$^{1}$ Department of Theoretical Physics, University of
Debrecen, P.O.\ Box 5, H-4010 Debrecen, Hungary \\
$^{2}$ Bay Zolt\'an Nonprofit Ltd.\ for Applied Research, H-1116 Budapest, Feh\'erv\'ari \'ut 
130. Hungary}

\begin{abstract}
We study the statistical features of magnetic noise accompanying the dynamic fracture of steel 
samples during mixed type fracture, where the overall ductile character of crack propagation
is interrupted by a sudden brittle jump.  
The structure of the voltage time series is investigated by identifying discrete peaks 
which correspond to elementary steps of the jerky cracking process. 
We show that the height, duration, area, and energy of peaks have power law distributions 
with exponents falling close to the corresponding values of pure ductile fracture. 
For single peaks a power law correlation of the height and area with the width is evidenced, however, 
the mixed nature of fracture gives rise 
to a crossover between two regimes of different exponents. The average pulse shape of micro-cracking 
events has a parabolic form with a right handed asymmetry similarly to quasi-static fracture propagation. 
The asymmetry emerges due to stress localization at the advancing crack front.
\end{abstract}
\maketitle

Dynamic fracture of heterogeneous materials plays a significant role in a broad
range of industrial applications from machine engineering to the safety of large scale 
constructions. For the assesment of the dynamic fracture strength of materials a detailed 
understanding of the initiation and propagation of cracks is essential 
\cite{freund_dynfrac_book}. 
Contrary to static fracture, the precise experimental characterization 
of dynamic fracture toughness is still a challenging problem.
One of the main issues of experimental techniques is the determination of the onset 
time of crack propagation which is typically done by crack-tip monitoring using high 
speed photography, single wire fracture gages, or strain gages 
\cite{freund_dynfrac_book,siewert_pendulum_2000}.
Due to the disorder present at the micro- or meso-scales of construction materials 
the propagation of cracks has been found to be a 
jerky process composed of discrete jumps which give rise to the emission 
of acoustic waves \cite{sethna_crackling_2001,alava_statistical_2006}. 
Dynamic fracture propagation in ferromagnetic materials like steel 
is also accompanied by magnetic emission (ME) which yields an important alternative
to acoustic testing \cite{siewert_pendulum_2000,FFE:FFE143}. 
Such crakling noise measurements 
can help to overcome the limitations of traditional techniques.
Recently, the integrated ME signal
has been suggested to be capable to detect fracture initiation, however, the reliability 
of the method is not satisfactory \cite{FFE:FFE143,siewert_pendulum_2000}. 
More information is encoded in the structure and evolution of the ME time series
of crackling events which can complement studies with integrated 
signals \cite{kun_structure_2004}. 
To work out well-established experimental techniques based on ME analysis
with a high reliability a comprehensive understanding of the statistics 
and dynamic origin of ME signals is needed \cite{kun_structure_2004}. 

In the present paper we investigate dynamic crack propagation by means of the magnetic 
emission technique focusing on the case of mixed fracture where the overall 
ductile character of crack advancement is interrupted by a sudden jump characteristic 
for brittle fracture. We extend the standard ME analysis of the integrated statistics 
of pulse quantities by studying the shape of single pulses and the correlations of 
pulse characteristics. We demonstrate that the mixed-mode fracture gives rise to a crossover
in the width-height or width-area correlation of crackling pulses, i.e.\ 
power law correlations are pointed out, however, different exponents are obtained
for small- and large-sized pulses, typical for ductile and brittle crack propagation, 
respectively.
\begin{figure}
\begin{center}
\epsfig{bbllx=0,bblly=0,bburx=380,bbury=220,
file=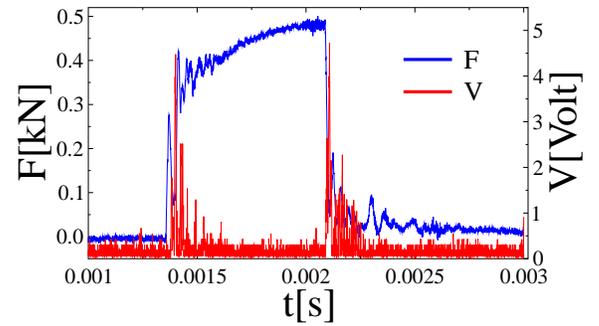, width=8.0cm}  
\caption{Force $F$ and voltage $V$ signal of magnetic emission measured during
the impact fracture of a V-notched JRQ RPV specimen. The overall ductile 
propagation of the crack is interrupted by a short brittle period indicated
by the sudden force drop accompanied by an intensive generation of high voltage pulses.
}
\label{fig:force_voltage}
\end{center}
\end{figure}
The average temporal profile of single bursts proved to have a parabolic 
shape with a right handed asymmetry which is the consequence of stress localization
at the propagating crack front.

In order to investigate dynamic fracture of steel, experiments 
were carried out by means of the Charpy impact machine 
\cite{francois_charpy_2002,FFE:FFE143,kun_structure_2004}.
The machine itself is a large size pendulum which is raised to a certain hight and hits
the specimen at the bottom of its swing. The specimens had the dimensions 
$1\mbox{cm}\times 1\mbox{cm}\times 5\mbox{cm}$ according to the standards
of Charpy tests \cite{francois_charpy_2002}. Steel specimens were
carefully manufactured in order to minimize the initial mechanical stress and damage 
inside the material. 
The velocity of the hammer of the impact machine
had the same value $v_0=5.5$m/s in all the experiments. 
In order to control the position of the crack initiated by the energetic hit
the specimens were either pre-cracked
or V-notched in the middle. Samples were made of two different types of steel, i.e.\
reactor pressure vessel steel {\it JRQ RPV} was used for both pre-cracked and V-notched specimens, while 
carbon steel {\it S355J2} was only used for V-notched samples.
Since the material of the samples has ferromagnetic properties the 
propagation of the crack is accompanied by changes of the magnetic field. The ME activity
was recorded in the form of a voltage signal by a small-sized coil attached to the impactor. 
Recently, it has been shown that this magnetic
noise provides interesting insight into the dynamics of crack propagation being sensitive 
to fluctuations of the speed of the crack front and of the opening of the crack 
\cite{FFE:FFE143,kun_structure_2004}.
 Parallel to magnetic noise we also measured the force acting 
on the moving impactor. The sampling rate was $\Delta t= 5\times 10^{-7}$s which provided 
sufficient resolution at our impact velocity $v_0$.

Figure \ref{fig:force_voltage} presents an example of the voltage signal $V(t)$ and of the
force $F(t)$ measured simultaneously during an impact fracture experiment. Note that
the type of fracture, i.e.\ either brittle or ductile can be inferred from the $F(t)$
curve: brittle fracture implies the sudden initiation of a crack running through
the sample which gives rise to a sudden drop-down of the force after a short uploading. 
The gradual evolution
of $F(t)$ observed in the figure is characteristic for ductile fracture where stable crack
propagation emerges. In our experiments all specimens had mixed-fracture, i.e.\ the crack
propagation started in a ductile manner, however, the sudden fall of the force indicates
a brittle jump of the crack which was then again followed by stable propagation (see Fig.\ 
\ref{fig:force_voltage}).
In spite of the complex cracking process, the force curve varies relatively smoothly, 
the wavy pattern is caused by elastic waves traveling through the sample. 
However, the voltage signal shows strong fluctuations with a complicated structure 
along the entire fracture process. Note that the brittle regime of fracture typically 
gives rise to a more intense magnetic activity with higher and broader peaks 
\cite{kun_structure_2004}.

\begin{figure}
\begin{center}
\epsfig{bbllx=40,bblly=15,bburx=440,bbury=300,
file=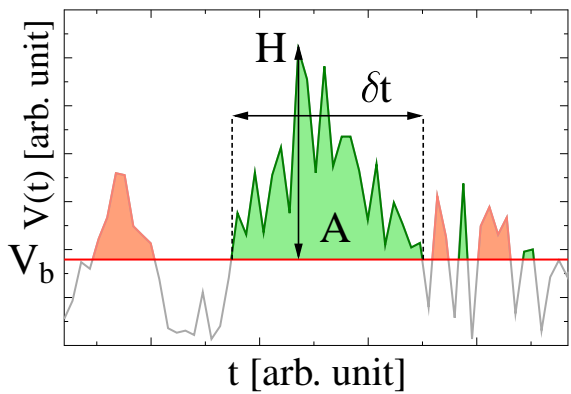, width=7.5cm}  
\caption{Identification of peaks and the definition of peak
quantities such as height $H$, duration $\delta t$, and area $A$. 
}
\label{fig:peak_illust}
\end{center}
\end{figure}
In order to characterize the structure of magnetic noise accompanying dynamic fracture 
we decompose the voltage time series into a sequence of discrete peaks following the method 
presented in Ref.\ \cite{kun_structure_2004}. Note that the method is analogous to the 
evaluation technique of Barkhausen noise measurements 
\cite{PhysRevE.54.2531,durin_scaling_2000,mehta_universal_2002,doi:10.1080/00018730802420614}.
As a first step we identify the background noise 
level $V_b$ by searching for the most probable voltage value. 
Then peaks are identified as segments of the $V(t)$ time series
between two intersections with the background $V_b$. 
In a single sample a few hundred voltage peaks are identified where peak $i$ starts 
at time $t_i^s$ and ends at $t_i^e$. In order to characterize single peaks a set of quantities
are determined by carefully evaluating the measured voltage signals, 
see Fig.\ \ref{fig:peak_illust} for illustration: The overall geometry
of peaks is characterized by their height $H_i$ and width, i.e.\ the peak duration $\delta t_i$.
The height $H_i$ of peak $i$ 
is the largest voltage value $V_i^{max}=\max \left[V(t)(t_i^s\leq t \leq t_i^e)\right]$ 
in the peak duration from which the background is subtracted
$H_i = V_i^{max} - V_b$.
The peak duration $\delta t_i$ is obtained as the difference of the ending $t_i^e$ and 
starting times $t_i^s$ as $\delta t_i = t_i^e - t_i^s$.
The area $A_i$ of peak $i$ is calculated as the integral of the voltage curve above the 
background level (see Fig.\ \ref{fig:peak_illust}), while for the energy of the peak 
$E_i$ the square of the voltage signal is integrated in a similar way
\beq{
A_i = \int_{t_i^s}^{t_i^e} (V(t)-V_b)dt, \qquad E_i = \int_{t_i^s}^{t_i^e} (V(t)-V_b)^2dt.
\label{eq:area_energy}
}

It has been shown in Ref.\ \cite{kun_structure_2004} that peaks of magnetic noise
can be associated to elementary events of dynamic crack propagation. 
Due to the stochastic nature of crack propagation all quantities defined above
have strong fluctuations, hence, for the characterization of the voltage time series
we determined the probability distributions and the correlations of peak quantities.
In order to improve the statistics of the data experiments were repeated 5, 7, and 12 times
for pre-cracked, V-notched JRQ RPV, and for V-notched carbon steel, respectively.
\begin{figure}
\begin{center}
\epsfig{bbllx=35,bblly=15,bburx=730,bbury=640,
file=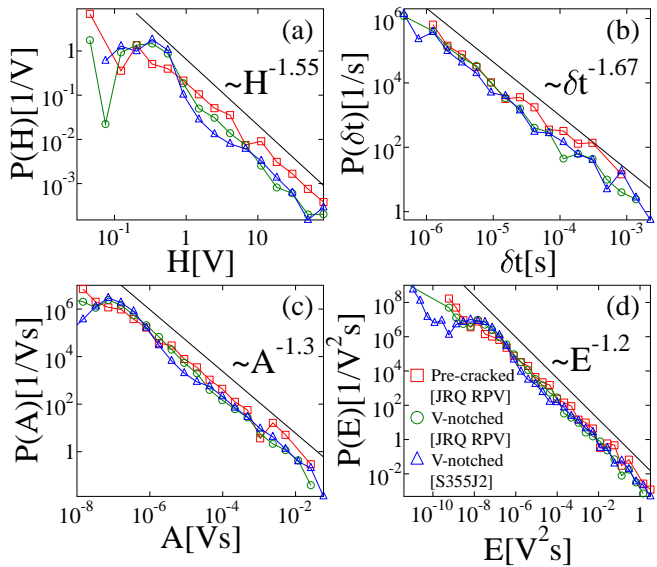, width=8.2cm}  
\caption{Probability distributions of the height $H$ $(a)$, duration $\delta t$ $(b)$,
area $A$ $(c)$, and energy $E$ $(d)$ of voltage peaks. Power law functional form is 
evidenced in all cases.
}
\label{fig:alldists}
\end{center}
\end{figure}

The primary quantities characterizing the shape of peaks are the width $\delta t$ and height $H$.
Figure \ref{fig:alldists}$(a)$ presents the probability density of peak heights $p(H)$
for all three sets of experiments. Power law decay of $p(H)$ can be observed for large 
values of $H$
\beq{
p(H) \propto H^{-\beta},
\label{eq:hdist}
}
where the exponent $\beta$ has nearly the same value $\beta = 1.55\pm 0.15$ for all 
types of samples.
The corresponding distribution of peak durations $p(\delta t)$ 
has also a power law asymptotics 
\beq{
p(\delta t) \propto \delta t^{-\alpha}
\label{eq:dtdist}
}
persisting over three decades in $\delta t$ (see Fig.\ \ref{fig:alldists}$(b)$).
The value of the exponent can be determined with a relatively good precision $\alpha=1.67\pm 0.06$.
Note that due to the low statistics of the data in the large $\delta t$ regime it is hard to
infer the shape of the cutoffs limiting the power laws.
The area $A$ and energy $E$ defined by Eqs.\ (\ref{eq:area_energy}) 
provide measures for the magnitude of crackling events represented by the peaks. 
The probability distributions of peak area $p(A)$ and energy $p(E)$ are presented 
in Figs.\ \ref{fig:alldists}$(c,d)$, where a power law functional form is evidenced for both
distributions over several orders of magnitude
\beq{
p(A) \propto  A^{-\tau}, \qquad p(E) \propto E^{-\epsilon}.
}
The exponents were obtained numerically as $\tau=1.3\pm 0.05$
and $\epsilon=1.2\pm 0.05$.
The power law character of single peak quantities implies that the jerky steps
of crack propagation are not random but correlations occur between consecutive jumps of the 
crack front. Ductile cracks advance in small steps as the hammer of the impact machine
proceeds. In case of brittle fracture the crack decouples from the impactor and it leads 
to sudden failure with a small number of large jumps \cite{kun_structure_2004}. 
Consequently, brittle and ductile cracking are 
characterized by lower and higher exponents of the distributions of peak quantities 
\cite{kun_structure_2004}.
A very interesting outcome of our measurements is that mixed type fracture is also characterized
by power law distributions which are naturally dominated by the peaks generated in the ductile 
regime.

Of course, single peak quantities are not independent of each other, i.e.\ it can be expected
that a broader peak has a larger height and in turn a larger magnitude represented by the peak 
area and energy.  
Such correlations of the peak height $H$ and duration $\delta t$ 
can be quantified by the conditional average, i.e.\ averaging $H$
for fixed values of $\delta t$.
 \begin{figure} 
 \begin{center}
 \epsfig{bbllx=50,bblly=25,bburx=730,bbury=330,
 file=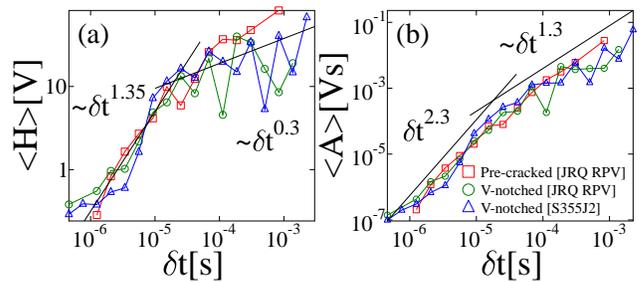, width=8.0cm}  
 \caption{Correlation of peak quantities. 
 Average height $\left<H \right>$  $(a)$ and average area $\left<A \right>$ $(b)$
 of peaks as a function of the peak duration 
 $\delta t$. Power law dependence is obtained with a crossover between two regimes of 
 different exponents. 
 }
 \label{fig:correl}
 \end{center}
 \end{figure}
Figure \ref{fig:correl}$(a)$ shows
that the average height $\left<H \right>$ has a power law dependence 
on the width of peaks $\delta t$
\beq{
\left<H \right> \propto \delta t^{\xi},
\label{eq:haverdt}
}
however, the value of the exponent $\xi$ has a distinct change at a characteristic 
peak duration $\delta t_c$. The crossover between the two power law regimes of different 
exponents occurs at $\delta t_c \approx 2 \times 10^{-5}$s and the exponent $\xi$ takes the 
values $\xi=1.35$ and $\xi=0.35$ below and above $\delta t_c$, respectively. 
The statistics of the data in Fig.\ \ref{fig:correl}$(a)$ is somewhat lower above the crossover 
point $\delta t >\delta t_c$ which is consistent with the rapid decrease of $p(\delta t)$
in Fig.\ \ref{fig:alldists}$(b)$. 
The result implies that 
for broad peaks $\delta t>\delta t_c$ the height increases slower with the width 
than for the narrow ones. 
Note that the crossover is not visible on the probability distributions $p(H)$ and $p(\delta t)$.
A possible explanation is that both the height and duration values above the crossover fall in the 
cutoff regime of the distributions. 

Based on Eq.\ (\ref{eq:haverdt}) we can assume the approximate 
relation $H\propto \delta t^{\xi}$ from which a scaling relation of the exponents 
$\alpha$ and $\beta$ of $p(\delta t)$ and $p(H)$ can be derived as
\beq{
\alpha = 1 + \xi(\beta - 1).
}
Substituting the value of the exponents the scaling form is fulfilled with a 
good precision below the crossover duration  but not above $\delta t_c$ since the 
exponents $\alpha$ and $\beta$ are dominated by the ductile regime of crack
propagation. Based on the height-width relation of pulses
the area $A$ can be simply approximated as a function of $\delta t$
\beq{
A\propto \delta t^{1+\xi}.
\label{eq:are_dt}
}
To test the validity of the above form we determined the conditional average 
$\left<A\right>(\delta t)$ by directly averaging $A$ in bins of $\delta t$.  
Figure \ref{fig:correl}$(b)$ demonstrates that power law behavior is obtained as expected
with the same crossover behavior that has been found for the peak height 
(see Fig.\ \ref{fig:correl}$(a)$).
The values of the exponent $2.3$ and $1.3$ below and above the crossover point are
consistent with Eq.\ (\ref{eq:are_dt}).

 \begin{figure}
 \begin{center}
 \epsfig{bbllx=30,bblly=30,bburx=380,bbury=330,
 file=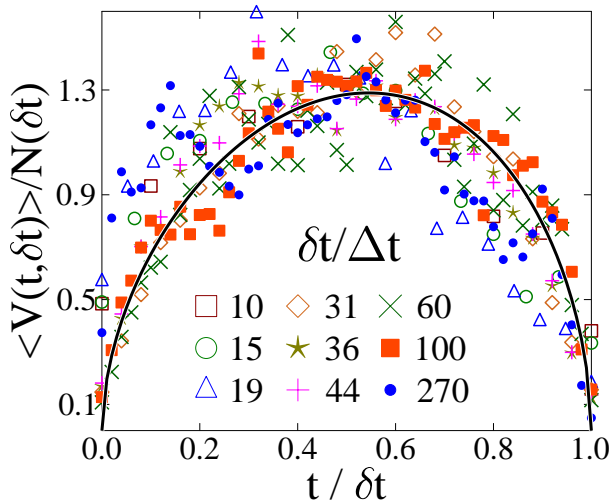, width=8.2cm}  
 \caption{Average temporal profile of pulses rescaled with $N(\delta t)$ varying the 
 duration $\delta t/\Delta t$ in a broad range from $10$ to $270$. 
 The continuous line represents best fit with Eq.\ (\ref{eq:scalingfunc}).
 \label{fig:pulseshape}}
 \end{center}
 \end{figure}
The correlation of the height, area, and duration provides a characterization of the 
overall shape of single cracking peaks. A more detailed description can be 
obtained by determining the average temporal profile $\left<V(t,\delta t)\right>$ 
of pulses of a fixed duration $\delta t$
\cite{doi:10.1080/00018730802420614,colaiori_shape_2004}. 
Since single voltage pulses correspond to individual steps of the advancing crack,
the average temporal profile provides information about the time evolution of 
intermittent crackling bursts. 
To improve the statistics, pulses of duration $\delta t\pm 0.1\delta t$ were averaged together
using the data of all measurements.
Figure \ref{fig:pulseshape} presents that rescaling $\left<V(t,\delta t)\right>$ 
with the norm $N(\delta t)=\int_0^1 \left<V(t/\delta t)\right>d(t/\delta t)$, 
pulse profiles of different duration $\delta t$ can be collapsed on a master curve 
as a function of the normalized time $t/\delta t$. 
It follows from Eq.\ (\ref{eq:are_dt}) that the norm scales as $N(\delta t) \sim \delta t^{\xi}$
so that the good quality data collapse implies the validity of the scaling form of
pulse shapes
\beq{
\left<V(t,\delta t)\right> = \delta t^{\xi}f(t/\delta t).
}
The scaling function $f(x)$ has a parabolic shape
which shows that during crackling jumps the crack 
front starts slowly then it accelerates
followed by a gradual slow down \cite{zapperi_nat_2012,danku_PhysRevLett.111.084302}. 
Note that the profile $f(x)$ is slightly skewed to the right indicating the time irreversible
nature of the temporal burst dynamics.
The same temporal profile has been found very recently for crackling bursts 
generated at a slowly propagating crack front in heterogeneous materials
\cite{danku_PhysRevLett.111.084302,laurson_evolution_2013}. 
It has been shown that the symmetry
of the pulse shape carries information about the range of stress redistribution
following local breaking events: long range interaction (mean field)
results in symmetric profiles, while short range load sharing gives rise
to a right handed asymmetry \cite{danku_PhysRevLett.111.084302,laurson_evolution_2013}. 
It follows that the right handed asymmetry of the pulse shapes in Fig.\ \ref{fig:pulseshape} 
is the fingerprint of the localized stress 
redistribution along the crack front during dynamic fracture propagation.
The scaling function $f(t/\delta t)$ was fitted with the functional form 
\begin{eqnarray}
f(x)\sim \left[x(1-x)\right]^{\gamma}\left[1-a(x-1/2)\right],
\label{eq:scalingfunc}
\end{eqnarray}
where the exponent $\gamma$ determines the shape of $f(x)$
in the vicinity of $x=0$ and $x=1$, while $a$ controls the degree of
asymmetry \cite{laurson_evolution_2013}.
Best fit was obtained with $\gamma=0.55$ which is significantly lower than the
corresponding mean field value $\gamma=1$ 
\cite{mehta_universal_2002,colaiori_shape_2004,zapperi_signature_2005,danku_PhysRevLett.111.084302}, 
however, it falls closer to the result on quasi-static planar 
crack propagation $\gamma \approx 0.78$ \cite{laurson_evolution_2013}. 
The asymmetry is quantified by the parameter value $a=-0.2$ which has a larger absolute value
than the ones predicted by models of static crack propagation 
\cite{laurson_evolution_2013,danku_PhysRevLett.111.084302}.
 
We studied dynamic fracture of steel by analysing the structure of magnetic 
emission time series generated by the jerky propagation and opening 
of a crack. 
Our analysis revealed that for all types of specimen 
the height, duration, area, and energy of crackling pulses 
are all power law distributed with exponents falling close to their counterparts
of pure ductile fracture.
The mixed (ductile-brittle) nature of the fracture
process gives rise to a crossover in the correlation of peak quantities
with a sharp transition between two regimes of different exponents.
The most remarkable result is that the average temporal profile of crackling 
peaks has a parabolic form with a right handed asymmetry similarly to pulse shapes characterizing
quasi-static crack propagation in heterogeneous materials.
Profiles of different durations can be collapsed on a universal master curve on both
sides of the crossover point. The asymmetry of pulse shapes 
can be attributed to stress localization at the propagating front.
The results imply that the time evolution of ME peaks is the same at the onset of crack propagation
as later in the regime of stable propagation since it is solely determined by the range of
stress redistribution. To obtain information on the onset time of fracture 
the analysis has to go beyond single peak quantities and involve e.g.\ the waiting
time between consecutive crackling events.

The work is supported by the projects No.\ TAMOP-4.2.2.A-11/1/KONV-2012-0036 
and OTKA K84157.

\bibliographystyle{/home/feri/papers/FRACTURE/GYONGYVER/REVISE/phaip}
\bibliography{/home/feri/papers/statphys_fracture}

\end{document}